# High power single crystal KTA optical parametric amplifier for efficient 1.4-3.5 µm mid-IR radiation generation

Bianka Csanaková[1,2], Ondřej Novák[1], Lukáš Roškot[1,2], Jiří Mužík[1], Martin Smrž[1], Helena Jelínková[2], Tomáš Mocek[1]

[1] HiLASE Centre, FZU - Institute of Physics of the Czech Academy of Sciences, Za Radnicí 828, Dolní Břežany, Czech Republic  
[2] Faculty of Nuclear Sciences and Physical Engineering, Czech Technical University in Prague, Břehová 78/7, Prague, Czech Republic

E-mail: ondrej.novak@hilase.cz

**Abstract**

In this paper, we present a single crystal, KTA (potassium titanyl-arsenate, $KTiOAsO_4$) based picosecond optical parametric amplifier pumped by an in-house built 1030 nm Yb:YAG thin-disk laser, capable of tunability from 1.46 to 3.5 µm, operating at 90 kHz, with high average power in the signal and idler beams. The highest output power of 8.9 W was reached for the 1750 nm signal beam with 19% conversion efficiency and the respective 2500 nm idler beam power was 6.2 W with 13% efficiency. The highest combined signal and idler mid-IR power was obtained at 17 W at the 2060 nm wavelength degeneracy point.

## 1. Introduction

In recent decades, the field of laser technology has witnessed remarkable advancements, leading to the development of laser sources emitting in the mid-infrared (mid-IR) spectral range. The mid-IR region, typically spanning wavelengths from approximately 2-10 µm [1], has gained significant attention for its unique properties and promising applications across various scientific and industrial domains. One particularly exciting development is the emergence of picosecond mid-IR laser sources, which offer a compelling combination of ultrafast pulse durations and the advantages of mid-IR radiation, such as reduced thermal damage, enhanced signal-to-noise ratios, and the ability to investigate ultrafast dynamic processes in various materials.

Unlike the near-infrared or visible regions, mid-IR radiation can interact with numerous molecular vibrations, providing access to a wide range of molecular species and enabling the investigation of molecular structures and dynamics with unprecedented precision [2]. The so-called "atmospheric window" and the molecular fingerprint region allow the detection of trace gases, making them invaluable in ensuring public safety and environmental preservation [3]. Moreover, due to the absorption peak of water at 3 µm, mid IR sources are invaluable in medicine, with lasers being routinely used in ophthalmology, dermatology, urology, and many other fields.

The farther range of the mid-IR spectrum, primarily characterized by strong absorption in diverse polymers, holds particular significance for industrial applications, making it the preferred wavelength range for tasks like plastic welding, cutting, drilling, and micromachining [4].

The optical parametric amplifier (OPA) presents a simple second-order nonlinear mechanism of generating high power laser radiation in the picosecond domain, where a high-power pump beam interacts with a seed beam in a nonlinear crystal, to amplify the seed beam (signal) and generate a new beam (idler). The seed beam typically stems from an optical parametric generator (OPG), or supercontinuum (SC).

In the picosecond regime the OPG-OPA scheme is widely recognized [5].
Several works utilizing KTA crystals as OPA media were published. In [6] a CEP stable mid-IR source is presented, with tunability from 1.3 to 4.5 µm, based on a 1.4 ps, 100 kHz Yb thin-disk pumped KTA crystal. Seeding is provided by supercontinuum generated in a YAG crystal. Total output power in signal and idler beams was 17 W. In [7] a 100 kHz, non-collinear KTA OPA is demonstrated. The seed is provided by white light generated in an uncoated YAG plate by a 100 kHz, 180 fs, 40 µJ pump provided by the Light Conversion Pharos system. The first OPA stage consists of a 1 mm thick MgO doped PPLN crystal seeded by the 1.4-1.7 µm portion of the white light. The second OPA stage is based on AR coated, 2 mm thick KTA crystal used in noncolinear geometry with angular dispersion compensation provided by a reflection grating. The gain reached a factor of 20 with pump-to-signal and pump-to-idler energy conversion efficiency over 27% and 12%, respectively. Overall central wavelength tunability spanned from 1.53 µm to 3.13 µm.

KTA used in both collinear and noncollinear geometry is used in [8] to obtain CEP stabilized 10-cycle optical pulses at 3.5 µm. Here, the pumping is provided by a Ti:Sapphire laser providing 5.5 mJ at 300 Hz. Seeding of the noncollinear KTA stage is provided by white-light generation in a YAG plate. The subsequent collinear KTA stage is then used to provide idler at 3.5 µm. Passing the 120 fs idler beam through Si plates and a $CaF_2$ dispersion compensator, the pulses are compressed down to 21 fs.

Other work reporting mid-infrared picosecond OPA not limited to KTA crystals includes OP-GaAs optical parametric oscillator [9], operating at 100 MHz, delivering 9.7 W total output power from signal and idler combined with wavelengths 3093 and 5598 nm for signal and idler, respectively, or in [10] a ZGP OPA pumped by a 2 µm, 100 kHz, 4 ps Ho:YLF laser, seeded by a 1 µm pumped double-pass MgO:PPLN OPG idler beam tunable from 2.5 to 4.1 µm is used in order to obtain wavelength tunability between 2.5 and 12 µm. The combined mid-IR energy accounts to about 1 µJ and corresponds to efficiency of 8%.

In this paper, we present a picosecond optical parametric amplifier (OPA) based on a single KTA crystal pumped by an in-house developed Yb:YAG thin-disk laser, operating at 90 kHz and seeded by a double-pass optical parametric generator (OPG) based on a periodically poled lithium-niobate crystal (PPLN). The KTA crystal was deemed suitable as it offers transparency at the pump wavelength at 1030 nm and has high transmission up to 3.5 µm. Tunability from 1.46-3.5 µm with high average power in the signal and idler beams. The highest output power of 8.9 W was reached for the 1750 nm with respective idler power of 6.2 W. The highest combined signal+idler mid-IR power was reached at 17 W at the 2060 nm wavelength degeneracy point.

## 2. The pump laser

The pumping is provided by an in-house developed 1030 nm Yb:YAG thin-disk CPA laser system operating at 90 W output power at 90 kHz with a pulse duration of 1.3 ps [11] and beam quality of $M^2 \sim 1.2$.

The pump pulse autocorrelation trace can be seen in Fig. 1, the output spectrum can be seen in Fig. 2 and the beam profile is shown in the Fig. 2 inset.

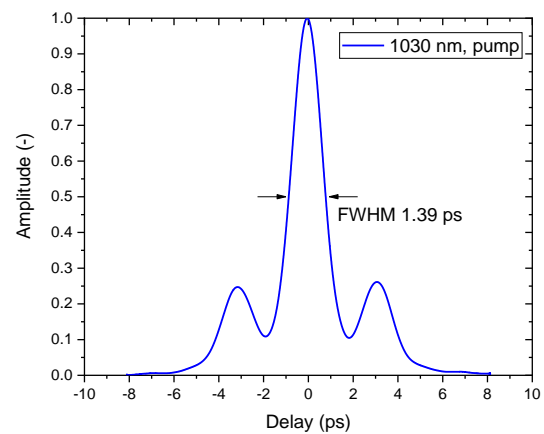

Fig. 1 Autocorrelation trace of the pump pulses.

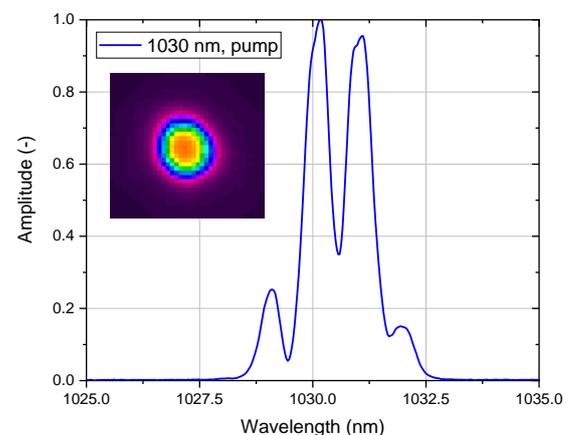

Fig. 2 Spectrum of the pump laser and its beam profile, shown in the inset of the figure.

### 3. Seed generator

In our case, the seed is provided by a double-pass optical parametric generator (OPG) based on a quasi-phase matched interaction in periodically poled lithium niobate (PPLN), crystal, described in more detail in [12]. The seed is based on signal beam stemming from the OPG stage. The signal beam is wavelength tunable between 1.46 and 2.06 µm. The latter wavelength corresponds to wavelength degeneracy. Signal output power was in the point range of 190 mW. Polarization of the seeding beam was horizontal, due to the *e-e-e* type 0 interaction taking place in the PPLN OPG, where *e* stands for extraordinary beam polarization.

The OPG delivered radiation in the range from 1459 to 2891 nm with up to 193 mW output power, with the wavelength degeneracy point at 2060 nm.

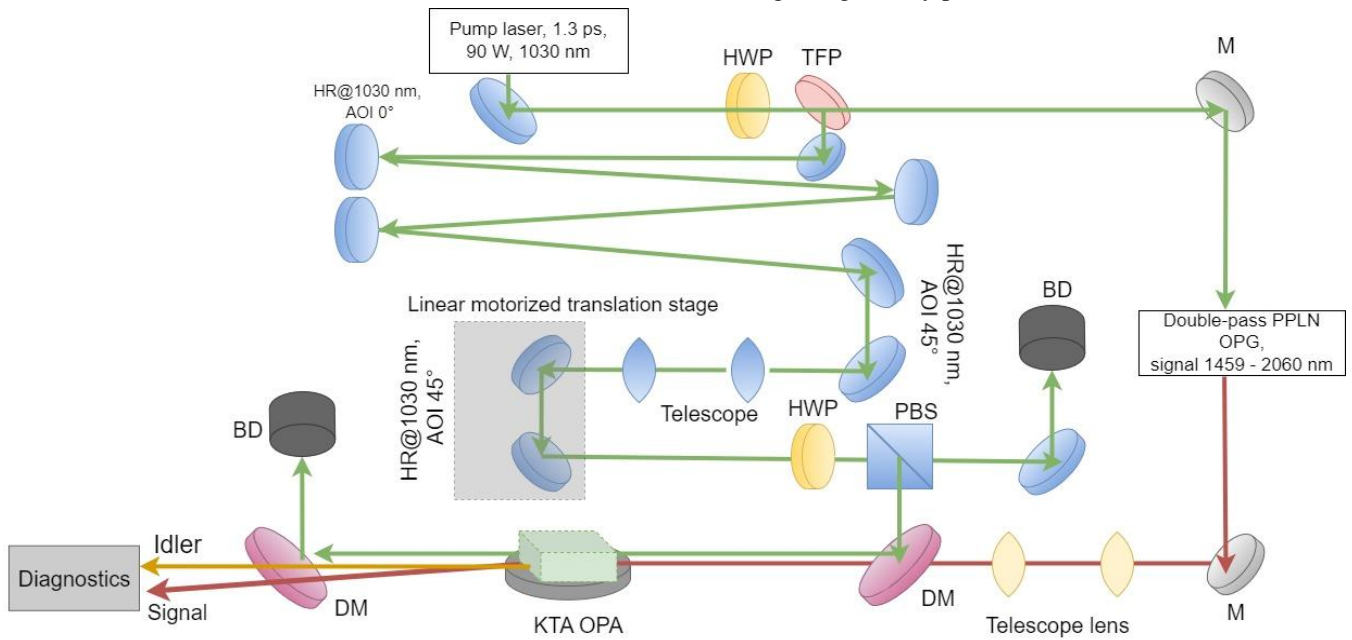

Fig. 3 OPA schematics. HWP – Half Wave-Plate, TFP – thin film polarizer, BD - Beam Dump, DM - Dichroic mirror (HR@1030 nm, HT@ 1400-2700 nm), PBS – polarizing beam splitter, M – silver coated mirror for seed radiation.

### 4. Optical parametric amplifier

We present a single crystal, KTA (potassium titanyl-arsenate, $KTiOAsO_4$) based optical parametric amplifier capable of tunability from 1.46 – 3.5 µm, operating at 90 kHz, with high average power in the signal and idler beams. The KTA crystal was chosen due to its high transmission in the whole expected tunability range [13], high ps damage threshold [14] and good commercial availability.

We present a single KTA crystal OPA, with focus on signal and idler beam tunability, output power, spectra, and beam profiles. Additionally, a characterization of the OPA output beams in terms of pulse length is presented. The schematic of the arrangement is presented in Fig. 3.

The OPA consists of a single KTA crystal used in type II interaction in the XZ plane (theta=44.85°, phi=0°), due to an approximately 75% higher effective nonlinear coefficient in comparison to the interaction in the YZ plane. The type II interaction of the *oeo (idler-signal-pump)*, where *e* represents the extraordinary and *o* the ordinary polarization, was chosen, due to its possibility of gap-free wavelength tuning and a lower required angular tuning range (around ± 4 ° internal angle as shown later in Fig. 5) as compared to the *eoo* interaction. Here we use the notation for extraordinary and ordinary beams, which is usually attributed to uniaxial crystals, even for the biaxial KTA in the specific plane of interaction XZ. The interaction type was near collinear.

The crystal was 10 mm long, with aperture 6×9 mm (height×width) placed on a rotation stage that allowed for horizontal tuning of the interaction angle. The crystal faces were AR coated (S1, S2: R<0.3% @ 1030 nm, R<1% @ (1500-2000 nm), R<2% @ (2000-3000 nm), R<4% @ (3000-3500 nm)). Pumping was provided by a part of the Yb thin-disk laser beam. A linear motorized delay stage was employed in the pumping branch for temporal overlap tuning of the interacting pump and seed beams inside the crystal, maximizing the generated output power. An attenuator consisting of a half-wave plate (HWP), and a polarizing beam sampler (PBS) was used to control the pumping power of the OPA, with 50 W maximum power used. The maximum pumping intensity on the crystal was 20 GW/cm$^2$. The polarization of the pumping beam was vertical. The seed was provided by the double-pass PPLN. The seeding beam size was adjusted by a pair of AR-coated $CaF_2$ lenses, to match that of the pumping beam in the crystal. Finally, the interacting beams were combined by the DM before the OPA stage and again that same DM separated the pump from the mid-IR beams at the exit of the OPA. The DM is highly reflective (R > 98%) at 1030 nm and highly transmissive in the 1.48 – 2.8 µm region (T > 80%). To deliver as much seeding power as possible, no bandpass filter was used in the seeding branch to separate the signal and idler beams generated by the OPG. Only the 1.46-2.06 µm portion of the OPG seeding is effectively amplified in the OPA, due to the KTA crystal cut preferring type II interaction between horizontally polarized signal and vertically polarized pump beams. Furthermore, the translation delay stage in the pumping branch ensures temporal overlap only with the signal pulse, as the idler pulse propagates with different group velocity through multiple optical elements, gaining different delay in respect to the seeding signal pulse. In addition, the seeding idler beam is highly divergent and is not reaching the KTA crystal effectively. Therefore, the only interaction taking place in the KTA crystal is between the 1030 nm pumping and the 1.46-2.06 µm seeding signal beam. The relationship between the generated wavelengths of the signal and idler beams is shown in Fig. 4. The internal angular wavelength tuning curve is presented in Fig. 5. The internal angle range is ~ ± 4 °. When accounted for the refraction on the air-KTA interface, the required external tuning angle range widens to ~ ± 7 °.

The length of the KTA crystal is limited by the group velocity mismatch between the pump and the amplified signal pulses. If the length of the crystal were to exceed the pulse-splitting length, the interacting pulses would no longer sufficiently temporally overlap, and the efficiency of the nonlinear interaction would decrease accordingly. In our case, the shortest pulse splitting length was calculated to be 22 mm for the signal wavelength of 1750 nm. However, to avoid self-focusing of the pumping beam, a 10 mm long crystal was chosen.

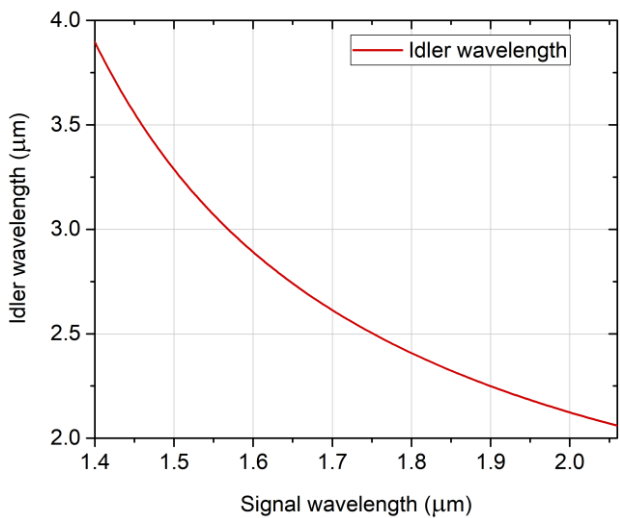

Fig. 4 Calculated relation between signal and idler wavelengths.

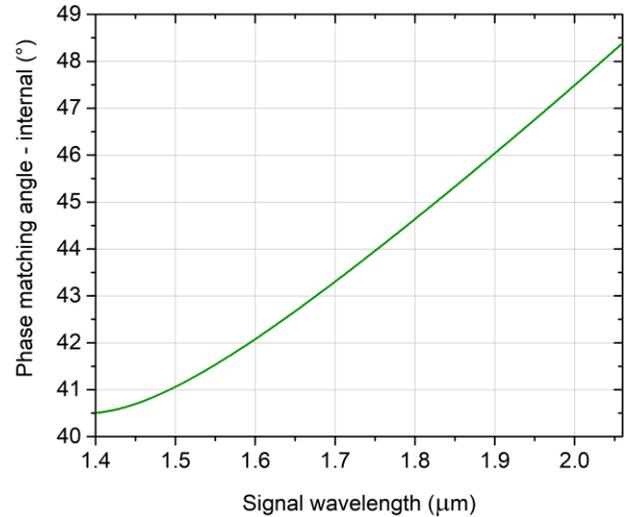

Fig. 5 Calculated internal phase matching angle dependence on signal wavelength.

### 3.1 Output power and beam profiles

The output power of the OPA and the pumping power were measured by the Ophir L50(150)A-BB-35 and Ophir 50(150)A-BB-26, respectively. The OPA power was measured behind the separating DM. Due to the nature of the interaction (type II), the signal and idler beam polarizations are orthogonal, therefore they can be separated by the Si plate polarizer, described in [15]. Due to the nature of the interaction, there is a spatial walk-off between the signal and idler beams, therefore the polarizer was adjusted for each separate beam due to its limited aperture.

The dependence of the generated average output power on average pump power are presented in Fig. 6. As can be seen from this figure, the highest output power of 8.9 W was reached for the 1750 nm signal beam. The respective idler beam at 2500 nm reached output power up to 6.2 W. The highest combined mid-IR power of 17 W was reached for the 2060 nm degeneracy point.

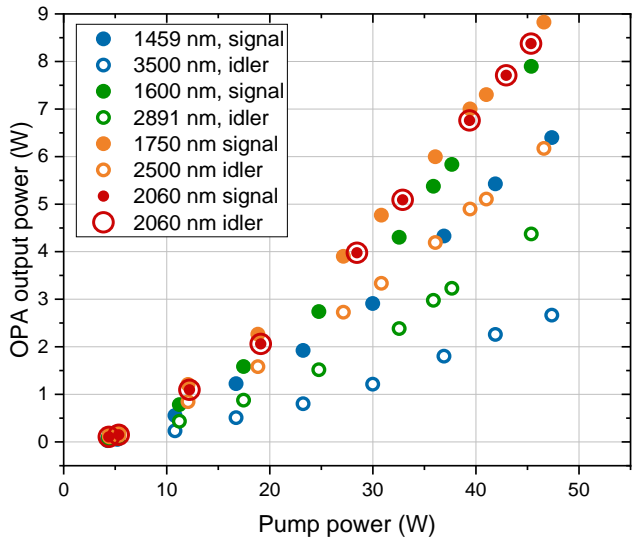

Fig. 6 Output power dependence of signal and idler beams on pump power.

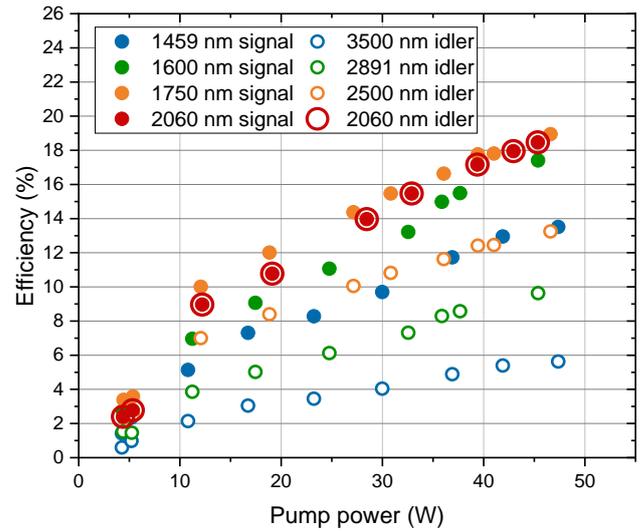

Fig. 7 Conversion efficiency of the KTA OPA in dependence on pump power.

The efficiency of the OPA interaction is depicted in Fig. 7. The highest conversion efficiency was reached for 1750 nm signal beam at 19%. The highest conversion efficiency for an idler beam was reached at 2060 nm at 18.4%. The highest photon conversion efficiency (signal + idler) was reached at the degeneracy point at 2060 nm at 34%. The signal + idler gain of the OPA stage was between 70 and 110 for the whole wavelength range. As can also be seen from Fig. 7, the efficiency for all wavelengths is approaching saturation behaviour with increasing pumping power.

This was confirmed by adding another 10 mm long KTA crystal in a walk-off compensation arrangement in addition to the first crystal. At 1750 nm and 2500 nm, the maximum output power from two crystals was 18.4 W signal + idler power, which is only a 3 W increase as compared to the single crystal stage at this wavelength combination. Shortly after this experiment, the second crystal was damaged due to self-focusing of the pumping beam in the crystal. It can therefore be said that increasing the crystal length of the OPA stage offers no additional benefits.

The output power stability of the generated signal and idler beams was also measured and is presented in Fig. 8. The pumping power stability was down to 0.5% and the seeding beam stability varied from 0.7% for 1750 nm to 1.1% for 1459 nm. As can be seen from Fig. 8, the power stabilities of generated signal and idler beams are below 2% for pumping powers above 10 W, apart from 3500 nm, which is slightly above 2% even pumping powers exceeding 10 W.

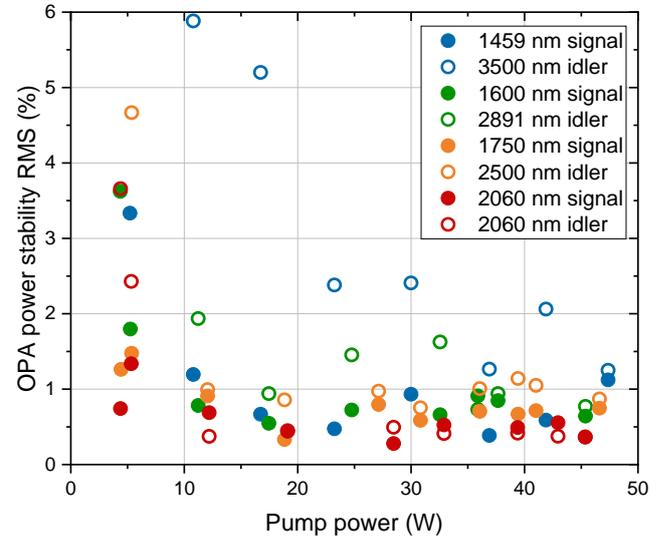

Fig. 8 Output power stability (RMS) dependence of signal and idler beams on pump power.

.

Signal and idler beam profiles (Fig. 9) at 1600 nm and 2891 nm, respectively, were captured by the Spiricon Pyrocam III. The beam separation was provided by the aforementioned silicon Brewster plate polarizer. However, due to the spatial walk-off between the beams and the different angle of refraction of both beams in the crystal, the polarizer was adjusted for the signal beam. Therefore, small clipping of the idler beam, which is passing through the polarizer at an angle, is visible in Fig. 9 b. The beam profiles are smooth and well-shaped.

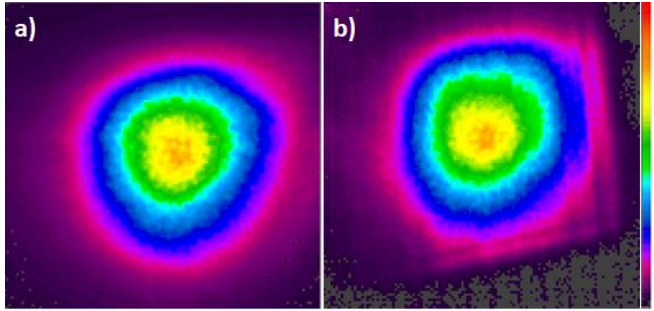

Fig. 9 Beam profiles of OPA radiation at 1600 nm signal (a) and 2891 nm idler (b).

### *3.2 KTA OPA spectra and tunability range*

The OPA output spectra are presented in Fig. 10. The tunability spans the range from 1459 nm to 3500 nm, the signal and idler beams up to 2400 nm were measured by the Optosky ATP8000 fiber-coupled spectrometer and idler beams beyond 2400 nm were measured by another spectrometer Andor Shamrock 303i with a pyroelectric line array detector PY-LA-S-510. The presented normalized spectra were not corrected for the spectral response of the spectrometer, which is supposed to be sufficiently flat for each measured narrow-band spectrum.

The wavelengths tunability range of the KTA OPA spans from 1459 nm to 3500 nm. The average spectral width is 40 nm FWHM. The OPA dramatically decreases the width of the seeding beam spectrum stemming from the OPG stage, resulting in narrower signal as well as idler beams as can be seen in Fig. 11. Other seeding beam spectra can be found in [12].

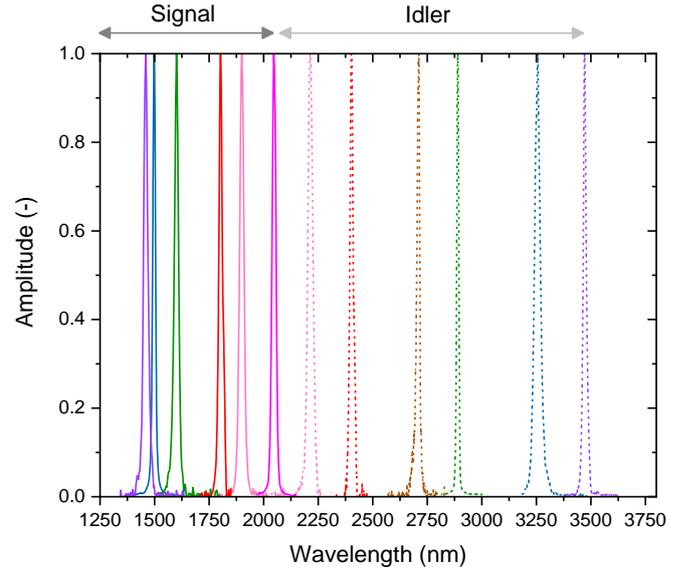

Fig. 10 Wavelength tunability of the KTA OPA. The solid lines present signal wavelengths, dashed lines indicate idler wavelengths. The color coding indicates signal-idler pairs.

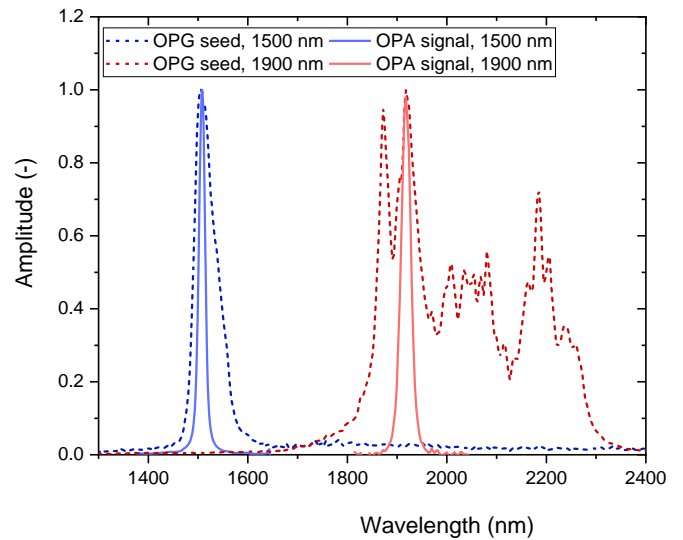

Fig. 11 Spectra comparison between OPG seeding beam (dashed) and generated OPA signal (solid).

### *3.3 Pulse length and autocorrelation*

For the KTA OPA, signal beam autocorrelation traces were measured by the APE PulseCheck 50 autocorrelator with appropriate crystal modules and a photomultiplier detector supplied with the autocorrelator. The autocorrelation trace for 2060 nm signal is presented in Fig. 12 for pumping power of 12 W. For higher pumping powers the temporal side lobes become more prominent, as can be seen in Fig. 13.

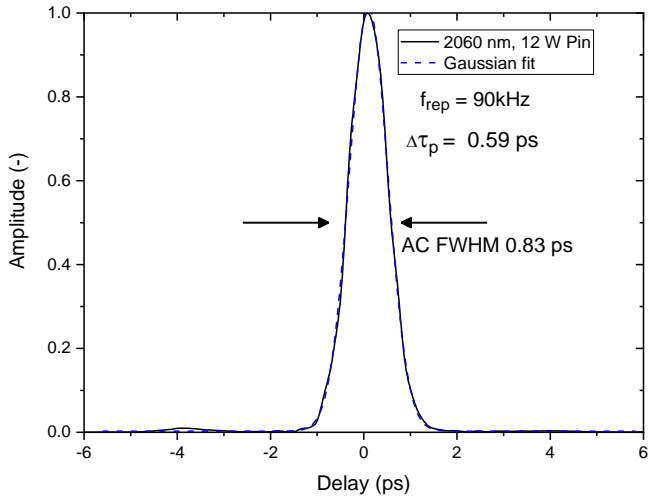

Fig. 12 Autocorrelation trace and gaussian fit for 2060 nm signal beam at 12 W pumping power.

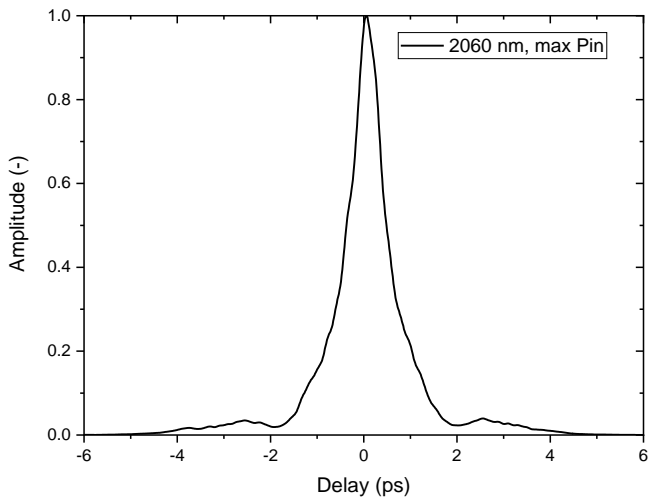

Fig. 13 Autocorrelation trace for 2060 nm signal beam at maximum pumping power.

For all wavelengths and the maximum pump power, the FWHM of the central peak of the autocorrelation trace is presented in Fig. 14. At 1459 nm the measured autocorrelation FWHM was 0.78 ps. With increasing wavelength, the AC trace FWHM exhibited an upward trend, with maximum of 1.23 ps at 1750 nm. With further increasing wavelength to 2060 nm, the measured ACF FWHM decreased, with a measured minimum of 0.89 ps at 2060 nm, thus all measured values being shorter than the pump pulse length of 1.39 ps. As seen from the measured values, the duration of the generated beams can become even shorter than the pump pulse itself, owing to the nonlinear dependence of gain on intensity. The edges of the pulse undergo less amplification than the central part, potentially resulting in a generated pulse shorter than the pump pulse. A simulation of the OPA stage under real pumping and seed conditions were made utilizing the SNLO software. Simulations include the effect of the group velocity mismatch between pump, signal, and idler pulses. Simulated signal pulse lengths are presented in Fig. 14 as blue data points. The simulated pulse lengths follow a very similar trend to values of the measured ACF FWHM.

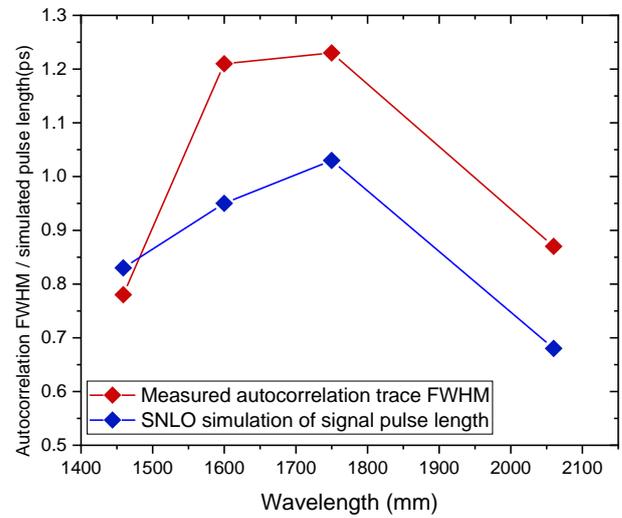

Fig. 14 Measured autocorrelation trace central peak FWHM, SNLO simulated pulse duration in dependence on the signal beam wavelength.

## 5. Conclusion

In this paper we presented a wavelength tunable, picosecond optical parametric amplifier (OPA) based on a single crystal KTA stage pumped by an in house developed ytterbium thin disk laser and seeded by a broadly tunable picosecond optical parametric generator (OPG) based on a periodically poled lithium niobate (PPLN) crystal.

This KTA OPA operating in the type II nonlinear interaction regime was characterized in terms of output power, efficiency, beam profile, wavelength tunability and pulse duration.

Maximum achievable mid-IR power was 17 W for the signal and idler combined output at degeneracy point of 2060 nm. Apart from the degeneracy point, maximum signal output power was 8.9 W at 1750 nm and maximum idler power was 6.2 W at 2500 nm. The highest conversion efficiency was reached for 1750 nm signal beam at 19%, for an idler beam at 2060 nm at 18.4%. The highest photon conversion efficiency (signal + idler) was reached at the degeneracy point at 2060 nm at 34%. The output fluence profiles showed no sign of beam deformation, remaining Gaussian like even for the highest pumping powers.

The output wavelength tunability spanned from 1459 nm to 3500 nm. This broad tunability offers to study laser-matter interaction in dependence on wavelength, e.g. dependence on the order of multi photon absorption. Pulse durations were estimated from the measured AC traces and showed that the estimated autocorrelation FWHM was lower than that of the pumping beam, with values up to 1.3 ps.

In conclusion, we have shown an efficient, broadly tunable high-power, single crystal mid-IR source based on a commercially available material, along with a simple double-pass OPG seeding stage, based on state-of-the-art pumping by an ytterbium thin disk laser technology at 1030 nm wavelength.

**Funding**

This work was co-funded by European Union and the state budget of the Czech Republic under the project LasApp CZ.02.01.01/00/22_008/0004573, and by the European Union's Horizon 2020 research and innovation programme under grant agreement No. 871124 (Laserlab-Europe).

**References**

[1] Hu J, Mawst L, Moss S, Petit L, and Ting D 2018 Feature issue introduction: Mid-infrared Optical Materials and their device applications *Optical Materials Express* **8** 2026

[2] Hochstrasser RM 2007 Two-dimensional spectroscopy at infrared and optical frequencies. *Proc Natl Acad Sci USA* **36** 14190-6.

[3] Ruizhi S, Peiheng Z, Wansen A, Yanning L, Ya L, Ruomei J, Wenxin L, Xiaolong W, Lei B, and Longjiang D 2019 Broadband switching of mid-infrared atmospheric windows by VO2-based thermal emitter. *Opt. Express* **27**, 11537-11546

[4] Bérubé JP, Frayssinous C, Lapointe J, Duval S, Fortin V and Vallée R 2019 Direct Inscription of on-surface waveguides in polymers using a mid-ir fiber laser. *Opt. Express* **27**, 31013-31022

[5] Manzoni C and Cerullo G 2016 Design criteria for ultrafast optical parametric amplifiers *Journal of Optics* **18** 103501

[6] Kanai T, Lee Y, Seo M, and Kim D 2019 Supercontinuum-seeded, carrier-envelope phase-stable, 4.5-W, 3.8-μm, 6-cycle, KTA optical parametric amplifier driven by a 1.4-ps Yb:YAG thin-disk amplifier for nonperturbative spectroscopy in solids. *J. Opt. Soc. Am. B* **36**, 2407-2413

[7] Heiner Z, Petrov V, Steinmeyer G, Vrakking M, and Mero M 2018 100-kHz, dual-beam OPA delivering high-quality, 5-cycle angular-dispersion-compensated mid-infrared idler pulses at 3.1 μm. *Opt. Express* **26**, 25793-25804

[8] Peiyu X, Faming L, Nobuhisa I, Teruto K, and Itatani J 2018 Generation of sub-two-cycle CEP-stable optical pulses at 3.5 μm by multiple-plate pulse compression for high-harmonic generation in crystals. *Opt. Lett.* 43, 2720-272

[9] Qiang F, Xu L, Liang S, Shardlow P, Shepherd D, Alam S, and Richardson D 2020 High-average-power picosecond mid-infrared OP-GaAs OPO. *Opt. Express* **28**, 5741-5748

[10] Hinkelmann M, Baudisch M, Wandt D, Morgner U, Zawilski K, Schunemann P, Neumann J, Rimke I, and Kracht D 2020 High-repetition rate, mid-infrared, picosecond pulse generation with μJ-energies based on OPG/OPA schemes in 2-μm-pumped ZnGeP$_2$. *Opt. Express* **28**, 21499-21508

[11] Novák O, Miura T, Smrž M, Chyla M, Nagisetty S, Mužík J, Linnemann J, Turčičová H, Jambunathan V, Slezák O, Sawicka-Chyla M, Pilař J, Bonora S, Divoký M, Měsíček J, Pranovich A, Sikocinski P, Huynh J, Severová P, Navrátil P, Vojna D, Horáčková L, Mann K, Lucianetti A, Endo A, Rostohar D and Mocek T 2015 Status of the high average power diode-pumped solid state laser development at Hilase. *Applied Sciences* **5** 637–65

[12] Csanaková B, Novák O, Roškot L, Mužík J, Cimrman M, Huynh J, Smrž M, Jelínková H and Mocek T 2023 Double-pass optical parametric generator pumped by Yb thin-disk laser for efficient 1.4–2.9 µm mid-IR radiation generation. *Laser Phys.* **33** 025005

[13] Hansson G, Karlsson H, Wang S, and Laurell F 2000 Transmission mea-surements in KTP and isomorphic compounds. *Appl. Opt.* **39**, 5058–5069

[14] Bach F, Mero M, Pasiskevicius V, Zukauskas A, and Petrov V 2017 High repetition rate, femtosecond and picosecond laser induced damage thresholds of Rb:KTiOPO4 at 1.03 μm. *Opt. Mater. Express* **7**, 744-750

[15] Csanaková B, Novák O, Smrž M, Huynh J, Jelínková H, Lucianetti A and Mocek T 2021 Silicon Brewster plate wavelength separator for a mid-IR optical parametric source. *Appl. Opt.* **60**, 281-290